# Growth of Tall Vertical Carbon Nanotube Forests using Al-Fe Catalysts and Transfer to Flexible Substrates

*Robinson Smith, Frank Brown, Ashley R. Chester*

Rio Salado College, Tempe, AZ 85281, USA

April 26, 2016

**Abstract**

*Carbon nanotubes (CNTs) have found plenty of applications in electronics, sensing, actuation, mechanical structures, etc. There is a growing demand to produce large scale arrays of CNTs that can satisfy the requirements of cost effective, quick and relatively simpler methods of growth. The other conflicting requirement is to produce very high quality of CNTs in very high densities and volumes. There has been considerable research that has gone into resolving the above mentioned issues. Here we show that a significantly simple and quick method of growing CNTs using Al-Fe thin film catalysts can also produce very dense and tall forests of vertically aligned CNTs that can be grown very uniformly over a substrate. We further show that this can be easily transferred onto a flexible substrate like an adhesive tape.*

Since the discovery of CNTs, there have been numerous applications and areas of research that have sprung up in several topics related to CNTs. Many applications are in the domain of electronics, sensing of physical parameters like pressure, energy storage, energy conversion, light trapping, actuation, field emission, etc., to name a few. Most of these applications require very different types of CNTs, like single walled carbon nanotubes (SWNTs), multi walled carbon nanotubes (MWNTs), etc., while they also require different conductivities, different physical attributes like length, etc. One thing that is common among all these requirements is simplicity of growth process, speed at which CNTs can be grown and the relative cost of growing them. There have been several efforts over the last two decades to understand the physical origins of different growth conditions and their influence on the resultant parameters of the CNTs. Several techniques have been developed to address many of the issues listed above, while some have focused on obtaining fast and simple growth. Here we address this issue by demonstrating a tall vertically aligned dense forest of CNTs grown using a simple thin film catalyst process. We demonstrate CNT forests with a height of 160 µm packed to a very high density. We discuss the process details that affect the growth, which include uniformity of the catalyst, contamination of the catalyst, gas flow conditions during growth and the treatment of the substrates prior to growth. We also demonstrate the ease at which these forests can be transferred onto a flexible substrate using the example of an adhesive fabric tape.

The process begins by the choice of a substrate, which in this case was a lightly boron-doped Si substrate that was 500 µm in thickness. The substrate was cleaned using the RCA cleaning process, which removes contaminants, ionic impurities, carbon-based impurities and residual oxides. The substrates were then baked in vacuum to drive away moisture. Following this, 5 nm of Al was deposited using e-beam evaporation and it was followed by deposition of 4.5 nm of Fe using e-beam deposition without breaking the vacuum in between (Figure 1a). This was followed by cleaning with acetone (heated to ~50 C) for 10 minutes, cleaning with isopropyl alcohol (heated to ~50 C) for 10 minutes and blown with dry nitrogen gas. The substrate was then made to undergo the catalytic CNT growth process, as described in Figure 1b. The first step in the process was to cleanse the substrate and also the tube furnace in which it was loaded using Ar gas (an inert gas) at room temperature (25 C) for 30 minutes, using Ar partial pressure of 1000 mTorr. This was followed by (step 2) ramping up of the temperature to 760 C over 15 minutes without changing the gas environment. Once the temperature was reached (step 3), the actual growth process was initiated. This included flow of ethylene gas at a partial pressure of 450 mTorr along with hydrogen gas at a partial pressure of 150 mTorr at 760 C. This step lasted for 6 minutes. Following this (step 4), the environment was changed to Ar flow at a partial pressure of 1000 mTorr at 760 C for 10 minutes. This was then followed by (step 5) a ramp down of temperature to room temperature, over >3 hours. Figure 1c is a schematic of the temperature profile during the various stages of the process. Figure 2 is a scanning electron microscope image of a typical CNT forest obtained using this process.

(a)

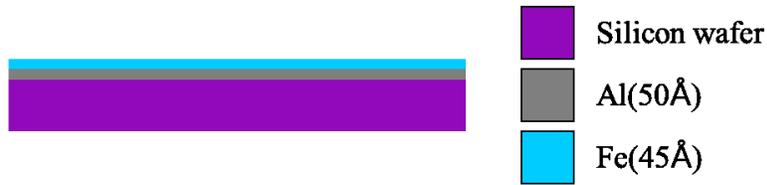

(b)

|  | Time(min.) | Channel 1 (Ethylene) | Channel 2 (Argon) | Channel 3 (Hydrogen) | Temp.(°C) |
|---|---|---|---|---|---|
| Step ① | 30 | 0 | 1000 | 0 | 25 |
| Step ② | 15 | 0 | 1000 | 0 | 760 |
| Step ③ | 6 | 450 | 0 | 150 | 760 |
| Step ④ | 10 | 0 | 1000 | 0 | 760 |
| Step ⑤ |  | 0 | 0 | 0 | 25 |

(c)

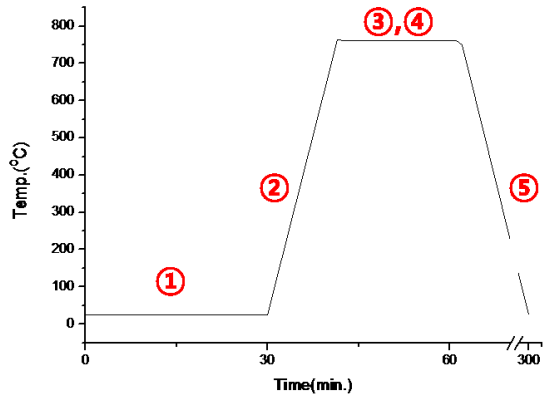

**Figure 1:** (a) Si substrate + 5 nm of Al was deposited using e-beam evaporation and it was followed by deposition of 4.5 nm of Fe using e-beam deposition. (b) Process flow at different steps labelled 1-5, described in the text. Numbers below gases in different channels are partial pressures in mTorr. (c) Temperature profile during the process depicted in (b).

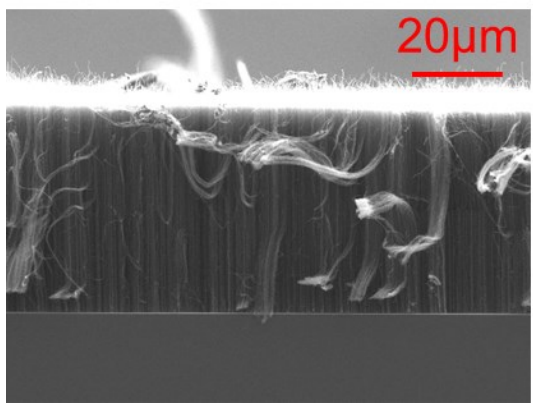

**Figure 2:** Side-view (cross-section) scanning electron microscope image of a typical CNT forest obtained using the technique described above.

In order to understand the effects and importance of sample preparation prior to the growth of CNTs, we made several similar samples undergo different preparations prior to CNT growth. These preparations were performed after the catalyst layers were deposited onto the Si substrate. Figure 3 summarizes the different sets of samples prepared for this experiment. The first set of samples underwent cutting into pieces appropriate for the tube furnace, while one underwent the standard cleaning process (acetone + isopropyl alcohol) and the other did not undergo the standard cleaning process. The second set of samples were heated on a hot-plate at 250 C for 10 mins to accelerate oxidation the Fe surface. In this set of samples, one underwent the standard cleaning process (acetone + isopropyl alcohol) while the other did not. The third set of samples were exposed to ambient atmosphere for 1 week to allow for a more natural pace of oxidation of the Fe surface. We then made each of these pieces individually undergo CNT growth, the results of which are shown in Figure 4.

## Preparation of various substrates

### Fe deposition

- Immediately processed
  - Cutting(20 × 20 mm)
    - Cleaning(Acetone → Alcohol) ①
    - Non-cleaned ②
- Heated on hotplate (250 °C for 10 min.) for oxidation
  - Cutting(20 × 20 mm)
    - Cleaning(Acetone → Alcohol) ③
    - Non-cleaned ④
- Exposed to ambient atmos. for 1 week for oxidation
  - Cutting(20 × 20 mm)
    - Cleaning(Acetone → Alcohol) ⑤
    - Non-cleaned ⑥

**Figure 3:** Chart of different samples labelled 1-6 with a description of the process they underwent.

Figure 4 displays CNT forests obtained from the growth process from the different samples (1-6) described above. The first obvious candidate in this matrix of images is sample 5, which had been exposed to ambient conditions for 1 week and had been cleaned. This displayed extremely poor growth of CNTs. This could be presumably because of the high extents of oxidation, moisture formation, etc. Sample 1, that underwent the cleaning process and was not exposed to environment or oxygen showed a height of forest of 50 μm. This was not the tallest among the lot.

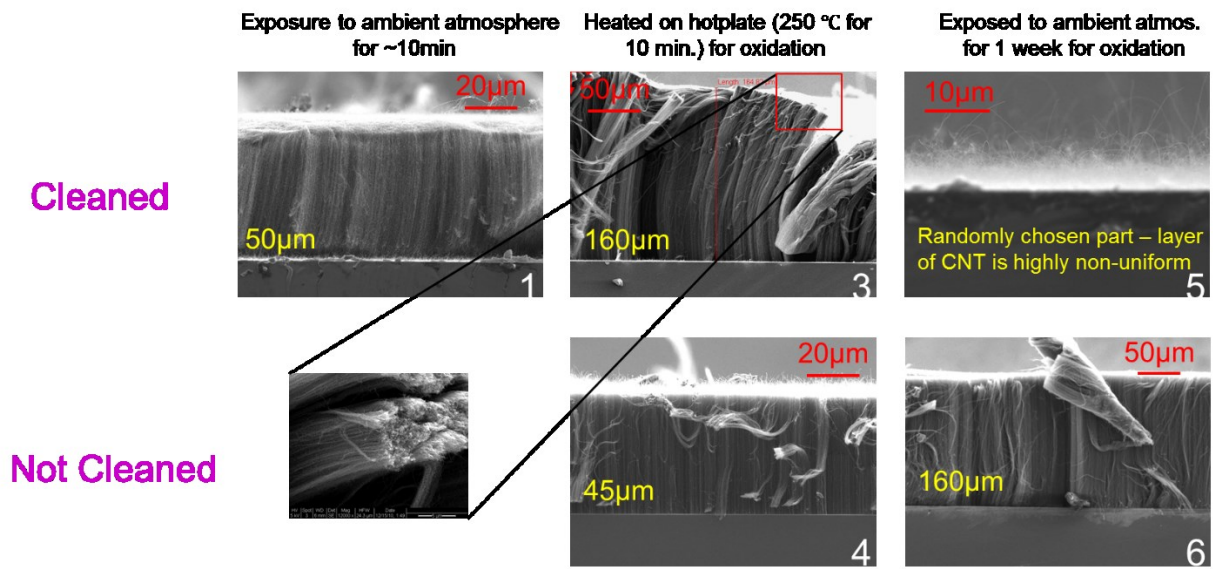

**Figure 4:** SEM images of different samples labelled 1-6, described in the text and in Figure 3. Red labels are length scales and yellow labels are the height of the forest of CNTs.

Samples 3 and 6 exhibited the tallest forests, with a height of 160 μm. Sample 3 had been heated on a hot plate in ambient conditions for 10 minutes and was also cleaned prior to growth of CNTs. This likely incorporated some moisture onto the surface, along with significant oxidation of the Fe catalyst. Sample 3 exhibited slightly bent CNTs and an apparently stressed forest. Whereas, sample 6 exhibited the best result of the lot, with the straightest and tallest CNTs. This sample was exposed to ambient conditions for a week and was not cleaned after that. This is a surprising result, compared to the cleaned version of a similar sample (sample 5). There could be many explanations for this result, but we believe it could be because of the incorporation of moisture that was loosely bound to the surface, which helped in the growth of CNTs. When the sample was cleaned (sample 5), the moisture was removed, and the other impurities remained, thereby proving detrimental for the growth of CNTs. On the other hand, not cleaning of the sample (sample 6) allowed for the moisture to stay, thereby greatly helping in the growth of CNTs. Not cleaning of sample 4 was likely hampering the growth of CNT by allowing the contaminants to remain on the surface, whereas cleaning the sample (sample 3) removed the contaminants, leaving behind only moisture, thereby aiding in CNT growth.

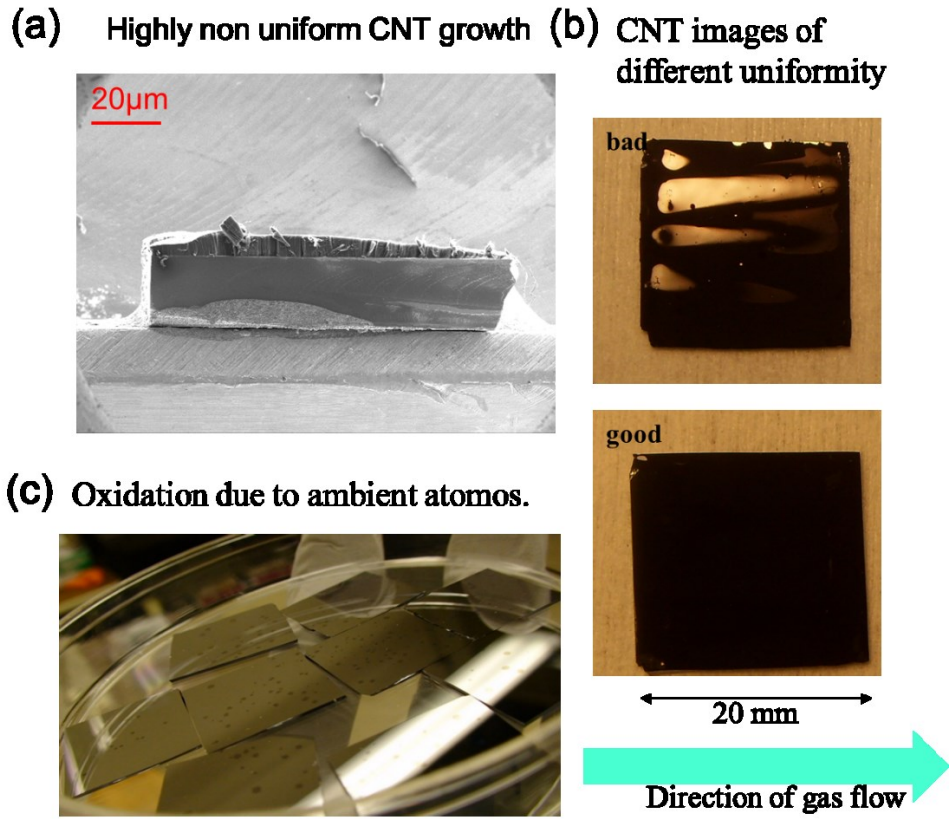

**Figure 5**: (a) Non-uniform CNT growth on sample 6 (referring to Figures 3-4). (b) Photographs of two different samples corresponding to non-uniform (bad) and uniform (good) CNT growth. Direction of gas flow is also indicated. Black regions are CNTs while shiny/bright regions are exposed metal. (c) Surface oxidation of samples treated similar to sample 6, displaying highly non-uniform discoloration corresponding to non-uniform surface oxidation.

However, there is one main cost of taking the approach as that in sample 6. Figure 5a shows a zoomed out SEM image of the sample, which shows clear non-uniformity in the growth of CNTs. Also, optical photographs of the sample prepared in a similar way to sample 6 ('bad' in Figure 5b) and of the sample prepared in a similar way to sample 1 ('good' in Figure 5b). There is a high degree of non-uniformity in sample 6 and similarly prepared samples. Figure 5c further illustrates this point and asserts that non-uniformity is a critical price to pay in this approach. We suggest that a better approach is to introduce a uniform, controlled level of moisture into the Fe layer such that there is no scope for agglomeration of the contaminants/moisture as seen in Figure 5c.

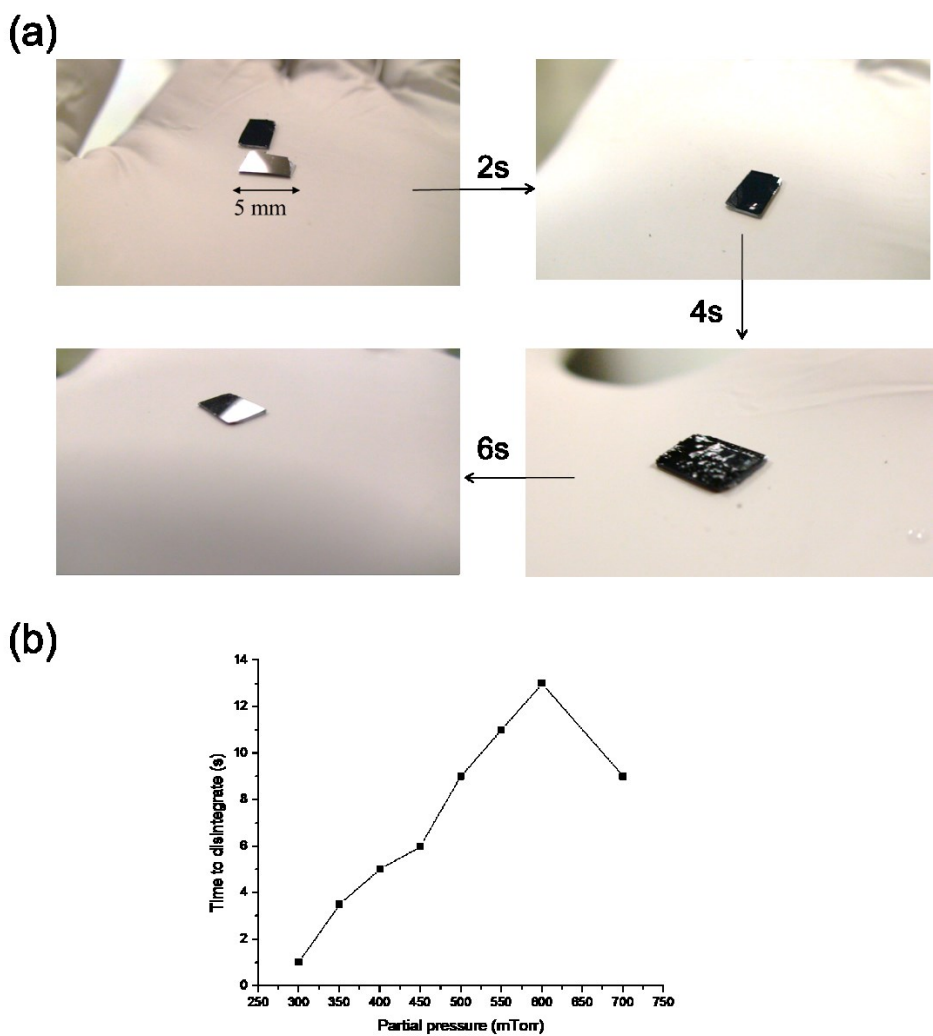

**Figure 6:** (a) Ultrasonication of a sample of CNT forest for different times (as noted). The first photograph is a comparison of a Si chip with (black) and without (shiny/bright/reflective) CNTs grown on it. After 6s, the sample has no CNTs left on the surface. (b) Time taken to completely

remove all CNTs from the surface of a sample for different partial pressures of ethylene during growth of CNTs.

In order to characterize the bonding strength of CNTs onto the surface of the substrate, we ultrasonicated the samples for different periods of times and observed to see if the CNTs had been peeled off (disintegrated). In the example in Figure 6a, sample 1 was used to discover that the CNTs would peel off after about 6s of sonication. We then varied the partial pressure of ethylene during the CNT growth process to see if that affected the time to disintegrate. The results are plotted in Figure 6b, where we see that the partial pressure of ethylene has a significant impact on the disintegration time. A partial pressure of 600 mTorr was found to provide the longest disintegration time of about 13 s, while increasing or decreasing the partial pressure of ethylene seemed to reduce the disintegration time.

We then attempted to transfer the layer of CNTs to a flexible substrate. We used a 3M electrical insulating fabric with adhesive and peeled the layer of CNTs off the Si substrate onto the adhesive tape (Figure 7). The tape was able to remove partially the forest/layer of CNTs on the substrate, while a part of it still stuck to the Si substrate. The density of this forest makes it convenient because despite imperfect transfer, there are still sufficient number of CNTs onto the transferred substrate that it is useful for many applications. In this experiment, we measured the resistance of the layer of transferred CNTs at its two ends and found that the resistance was about 4 k$\Omega$, showing that there is significant conductivity and also coverage of the CNTs. Further experiments for efficient transfer without using an adhesive layer are underway.

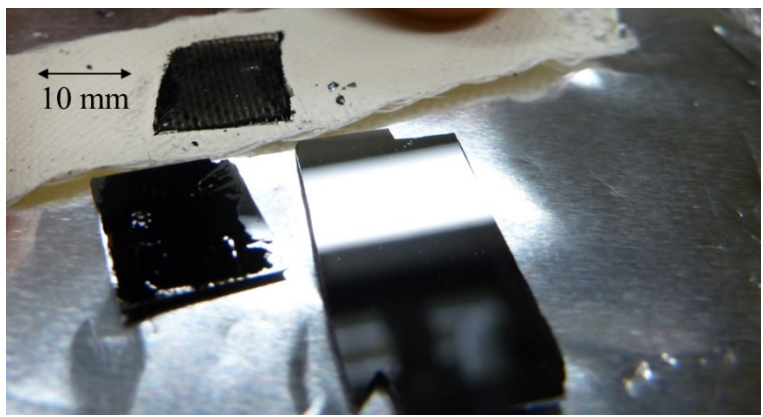

**Figure 7:** A layer of CNTs has been partially transferred onto an adhesive fiber tape, while a part of it still remains on the Si substrate of origin. Also shown in the photograph is a Si substrate without a layer of CNTs on it, for comparison.

In conclusion, efficient growth of CNTs is a very important area of research and addresses many utilities. Here we showed that the Al/Fe catalyst method produced tall and dense forests of CNTs while we also discussed optimization of growth and sample preparation parameters. We further showed that it was possible to transfer the forest of CNTs onto a flexible substrate using adhesives. This work is a step further in understanding different growth techniques meant for a variety of applications.